\documentclass[a4paper,11pt]{article}
\usepackage{pos}
\usepackage{float}
\usepackage{subcaption}

\usepackage{bm}
\def\qt{\bm q}
\def\kt{\bm k}

\def\mut{\bm \mu^\prime}
\def\mutsq{\mu^{\prime2}}
\def\nn{\nonumber}

\title{Recent developments in the Parton Branching approach to TMDs}
\author*[a]{Lissa Keersmaekers}
\author[b]{Sara Taheri Monfared}

\affiliation[a]{Department of Physics, University of Antwerp,\\
Groenenborgerlaan 171, 2020 Antwerpen, Belgium}
\affiliation[b]{Deutsches Elektronen-Synchrotron DESY,\\
Notkestra{\ss}e 85, D-22607 Hamburg, Germany}
\emailAdd{lissa.keersaekers@uantwerpen.be}
\emailAdd{sara.taheri.monfared@desy.de}

\abstract{The Parton Branching (PB) Method describes the evolution of transverse momentum dependent parton densities (TMDs). The obtained TMDs can be used in Monte Carlo generators to describe physical observables. We give an overview of recent developments, in particular the new application of PB TMDs in Drell-Yan + jets  production, the four-flavor and five-flavor schemes for heavy quark contributions, and the inclusion of QED corrections in the PB evolution.}

\FullConference{%
  *** The European Physical Society Conference on High Energy Physics (EPS-HEP2021), ***\\
  *** 26-30 July 2021 ***\\
  *** Online conference, jointly organized by Universität Hamburg and the research center DESY ***
}

\begin{document}
\maketitle

\section{Introduction}
The development of transverse momentum dependent (TMD) factorization theorems (\cite{Angeles-Martinez:2015sea} and references therein) has lead to an increase of precision in predictions of observables such as the Drell-Yan (DY) transverse momentum spectrum.\\
The Parton Branching (PB) method~\cite{HAUTMANN2017446,Hautmann2018,PhysRevD.99.074008} presents an angular ordered evolution for TMD parton distribution functions (TMD PDFs or TMDs), expressed in terms of real-emission splitting functions and Sudakov form factors.
The PB TMDs were fitted~\cite{PhysRevD.99.074008} to the full HERAI+II inclusive DIS data using the {\sc xFitter}~\cite{Alekhin2015} framework and are available in TMDlib~\cite{Hautmann_2014}, a library for TMDs and unintegrated PDFs. These TMDs were applied to Drell Yan production \cite{PhysRevD.100.074027,HAUTMANN2019114795,Martinez2020}.\\
With new software developments, such as the release of TMDlib2~\cite{Abdulov_2021}, which includes new functionalities (e.g. the treatment of TMD uncertainties), and the newest version of the Monte Carlo event generator {\sc CASCADE}~\cite{Jung2010} ({\sc CASCADE3}~\cite{Baranov2021}), which includes an intial state parton shower that is fully consistent with the PB TMDs, the applications of PB TMDs will increase.\\
This article will give an overview of recent developments within the Parton Branching method, both in terms of new applications as new developments in the evolution.

\section{Parton Branching Evolution equations}\label{sec:PBeq}
The Parton Branching evolution equations are given by:
\begin{align}
\tilde{\mathcal{A}}_a(x,\kt,\mu^2)=&{\Delta_a(\mu^2)}\tilde{\mathcal{A}}_a(x,\kt,\mu_0^2)\nn+\sum_b\int\frac{d^2\mut}{\pi\mutsq}\frac{\Delta_a(\mu^2)}{\Delta_a(\mutsq)}\Theta(\mu^2-\mutsq)\Theta(\mutsq-\mu_0^2)\times\\
&\times\int_x^{z_M}dz{P_{ab}(z)}\tilde{\mathcal{A}}_b(\frac{x}{z},\kt+(1-z)\mut,\mutsq), 
\label{eq:evolution}
\end{align}
with $\tilde{\mathcal{A}}_a(x,\kt,\mu^2)=x{\mathcal{A}}_a(x,\kt,\mu^2)$ the momentum weighted TMD of flavor $a$, with longitudinal momentum fraction of the proton $x$ and $\kt$ the transverse momentum, evaluated at scale $\mu$, $P_{ab}(z)$ the real-emission part of the DGLAP splitting functions for a splitting of parton $b$ to $a$, with $z$ the longitudinal momentum fraction and the Sudakov form factor for a parton of flavor $a$ is given by  $\Delta_a(\mu^2)=\exp[-\sum_b\int_{\mu_0^2}^{\mu^2}\frac{d\mu^{\prime 2}}{\mu^{\prime 2}}\int_0^{z_M}dz\ z\  P^{col}_{ba}(z,\mu^{\prime 2})]$. Angular ordering can enter the evolution through three aspects: i) the relation between the evolution scale $\mu'$ and the transverse momentum of the emitted parton $\qt$: $(1-z)\mu'=|\qt|$, which is embodied in all PB TMDs, ii) the scale of the strong coupling $\alpha_s(\qt^2)$ which is present in the fitted PB-NLO-HERAI+II-2018-set2, but not in PB-NLO-HERAI+II-2018-set1, which uses $\alpha_s(\mu'^2)$, iii) the dynamical (i.e. dependent on the evolution scale) soft-gluon resolution scale $z_M=1-q_0/\mu'$, with $q_0$ the minimal transverse momentum of the emitted parton. The resolution scale seperates resolvable from non-resolvable branchings. The effects of the dynamical resolution scale have been studied in \cite{HAUTMANN2019114795}.%, but have not yet been incorporated in the published fits which use a fixed resoltion scale. A natural development of the PB method is thus to perform fits that include the dynamical resolution scale, which are now being performed \cite{olaetal-inprep}.
\section{Multijet-merging}
\begin{figure}
\begin{minipage}{0.5\linewidth}
\centerline{\includegraphics[width=0.99\linewidth]{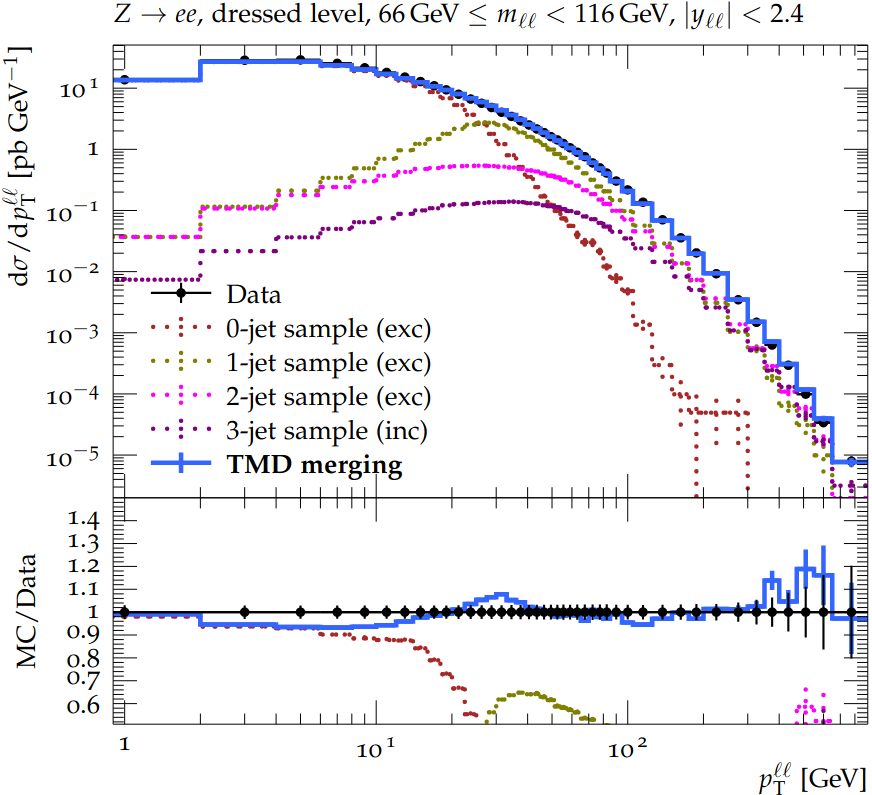}}
%\subcaption{$p_\bot$-spectrum}\label{fig:Zjets}.
\end{minipage}
\begin{minipage}{0.5\linewidth}
\centerline{\includegraphics[width=0.99\linewidth]{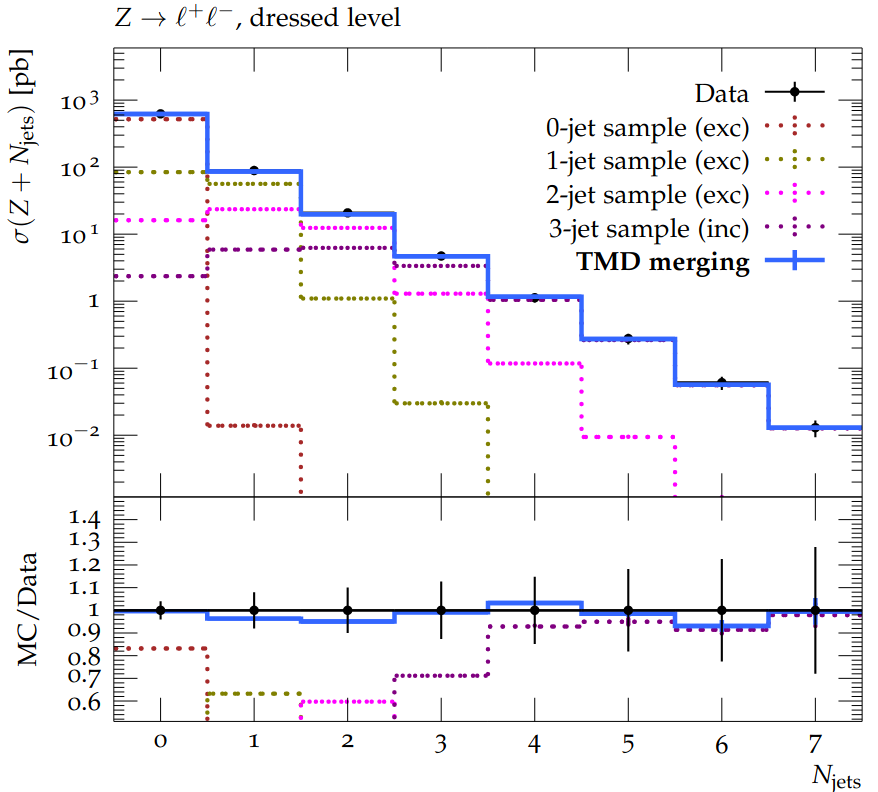}}
%\subcaption{Jet multiplicity}
\end{minipage}
\caption[]{Predictions obtained with TMD merging for the production of a Z-boson in association with jets. Left: Z-boson $p_\bot$-spectrum; Right: Jet multiplicity.  Figures from \cite{Martinez:2021chk}.}
\label{fig:multijet}
\end{figure}

Studies of TMD effects have been so far mostly used on low $p_\bot$-spectra of inclusive observables. However, the authors of \cite{Martinez:2021chk} realized that the large transverse momentum tails of TMDs, that arise naturally due to the renormalization-group evolution are used to describe multi-jet final states. A new "TMD merging" algorithm has been developed in \cite{Martinez:2021chk}, which extends the "MLM merging" procedure to include TMD initial state evolution. Compared to standard MLM, this method reduces systematical uncertainties and improves the description of higher-order emissions beyond the maximum parton multiplicity of the matrix element calculations.\\
In figure \ref{fig:multijet}, their prediction of the Z-boson $p_\bot$-spectrum and jet-multiplicity is shown. For these results the PB-NLO-HERAI+II-2018-set2 TMDs where used. The TMD merging algorithm describes the whole Z-boson $p_\bot$-range very well. The description of Jet-Multiplicity is remarkable, especially for multiplicities that are higher than the jet multiplicity of the matrix element, which is three. One can expect that the effects studied in their work will become even more important at future collider experiments, since TMD broadening grows with the evolution scale.\\
%\section{Leading Order fits}
%The TMD merging algorithm from the previous section is developed at Leading Order (LO) in the strong coupling. In contrast, the published fits were obtained with Next to LO (NLO) splitting functions. For consistency reasons, applications such as TMD merging could benefit from LO fits, which are being performed ~\cite{saraetal-inprep}.
%
\section{Photon TMD}
To obtain the same accuracy as current experimental programs, electroweak corrections should be applied to the before purely QCD evolution of the PB method. The most notable change
in the QED corrected evolution of parton distributions is the presence of the photon density. We determined both collinear and TMD photon densities with PB method ~\cite{Jung:2021mox}. 

\begin{figure}
\begin{minipage}{0.5\linewidth}
\centerline{\includegraphics[width=0.99\linewidth]{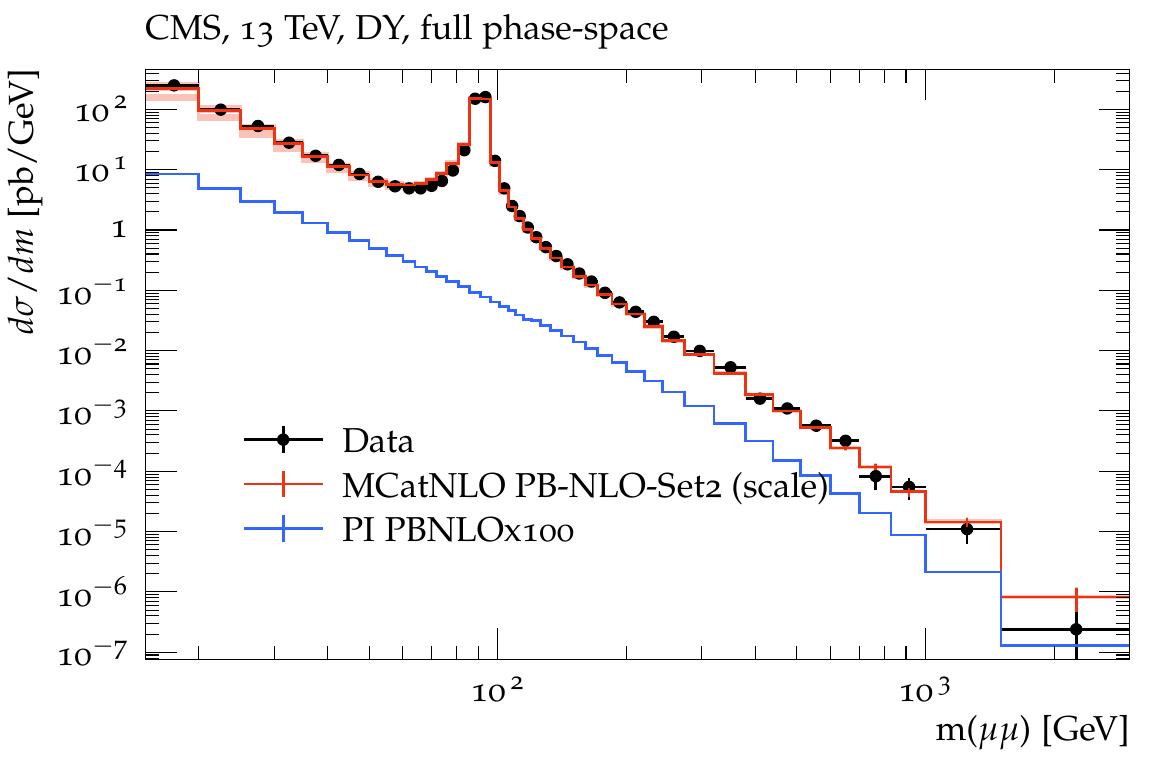}}
\caption*{a}
\end{minipage}
\hfill
\begin{minipage}{0.5\linewidth}
\centerline{\includegraphics[width=0.99\linewidth]{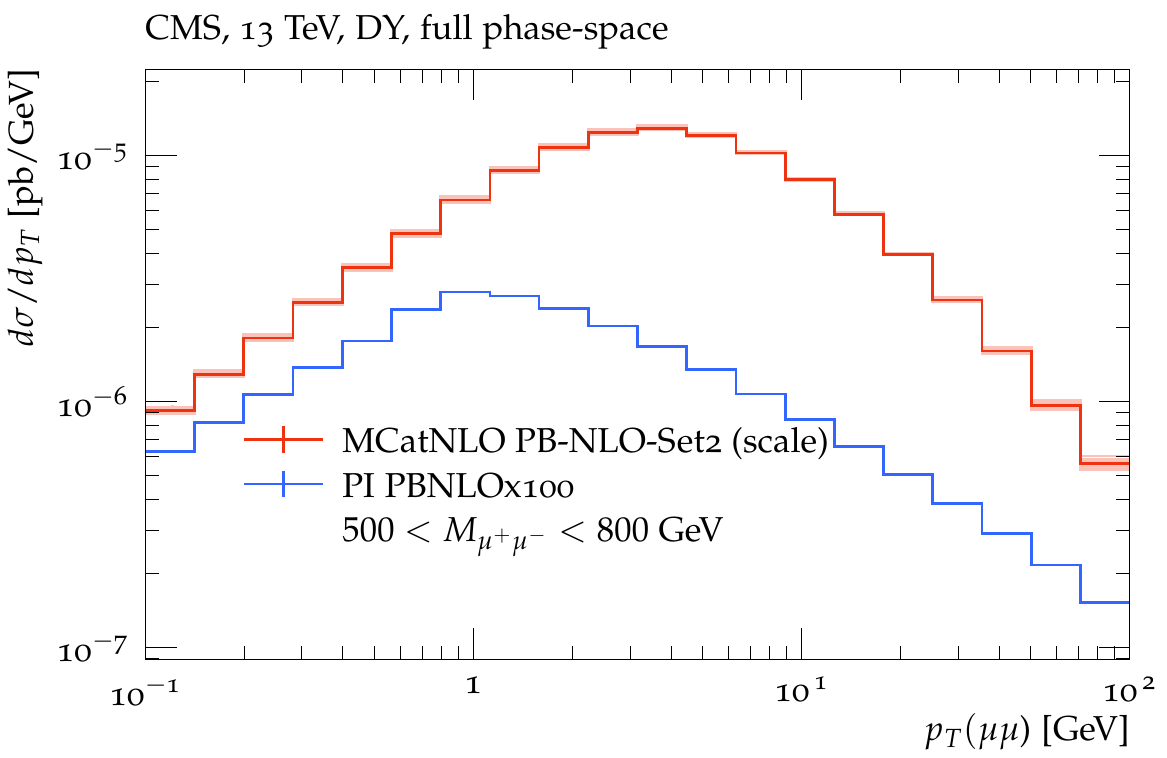}}
\caption*{b}
\end{minipage}
\caption[]{Standard DY and photon induced mass distribution (a) and transverse momentum spectra (b) based on collinear and TMD QED PDFs}
\label{fig:photon}
\end{figure}

At high mass Drell-Yan (DY) production, contributions from photon-photon scattering into lepton pairs play a role. The collinear NLO QED PDFs describe well the measured dilepton mass spectrum at LHC center-of-mass energies \cite{CMS:2018mdl}. As shown in Fig.\ref{fig:photon}.a, the small contribution from Photon-initiated (PI) lepton production is also determined. The photon TMD has been used to predict the transverse momentum spectrum of DY lepton-pair production at very high masses (Fig \ref{fig:photon}.b). 
 
\section{Four- and Five-flavor schemes}
\begin{figure}
\begin{minipage}{0.5\linewidth}
\centerline{\includegraphics[width=0.99\linewidth]{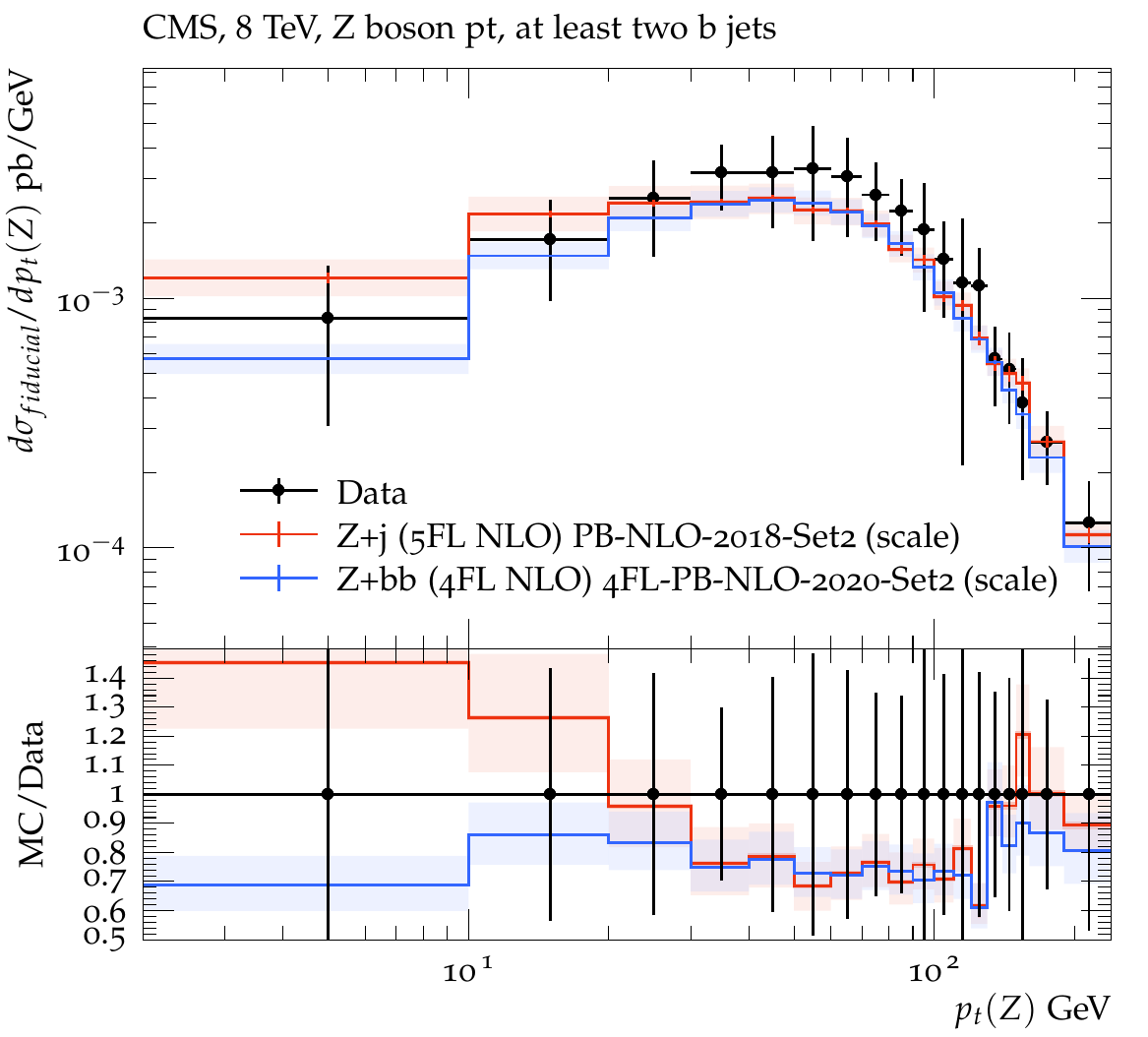}}
\caption*{a}
\end{minipage}
\hfill
\begin{minipage}{0.5\linewidth}
\centerline{\includegraphics[width=0.99\linewidth]{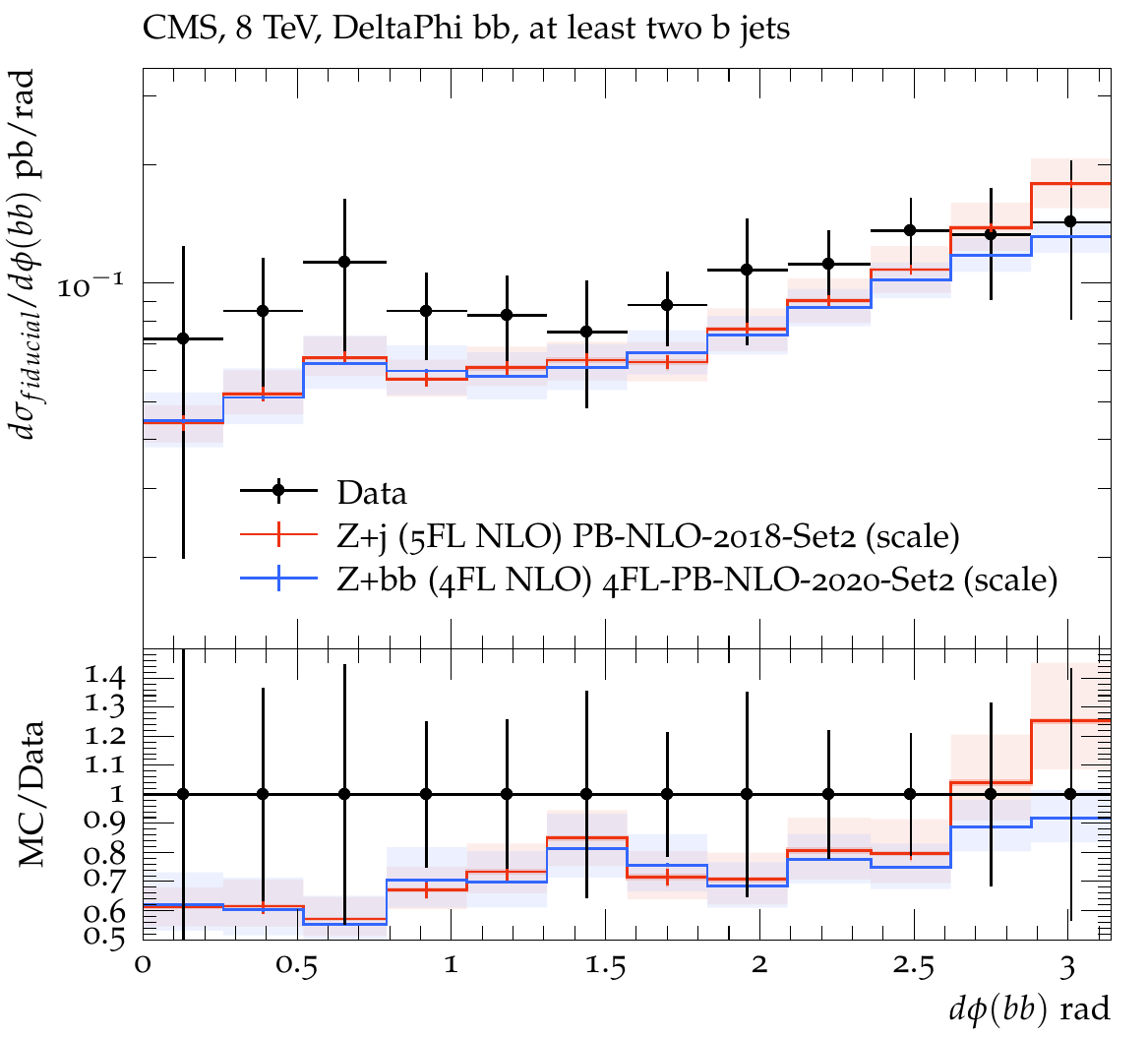}}
\caption*{b}
\end{minipage}
\caption[]{Differential cross section for $Z+\ge2 b$ jets production as a function transverse momentum of the $Z$ boson $p_t$ (a) and the azimuthal angular separation $\Delta\phi_{bb}$ between the directions of the two $b$ jets in the transverse plane (b). Shown are the predictions obtained in the 4FLVN- and 5FLVN- schemes.}
\label{fig:4FL-5FL}
\end{figure}

The first set of NLO collinear and TMD parton densities in four-flavor-variable-number (4FLN) scheme within the PB approach is determined \cite{Jung:2021vym}. The 4FLVN and five-flavor-variable-number (5FLVN) PB-TMD distributions \cite{PhysRevD.99.074008}
 were applied to predict $Z + b\bar{b}$ tagged jet production at LHC energies. In Fig. \ref{fig:4FL-5FL}, we show the predictions obtained within both schemes, which are in very good agreement with the measurements. 

The completely different configurations of heavy flavor collinear and TMD PDFs and the corresponding initial TMD parton shower in the 4FLVN and 5FLVN schemes allow 
for a precise investigation of the evolution of the PB-TMD PDFs as well as the PB-TMD parton shower.

\section{Conclusion}
The Parton Branching method has already had several successes last years, especially in the description of the low-$p_\bot$ spectrum of the Drell Yan process at both high and low energies\cite{PhysRevD.100.074027,HAUTMANN2019114795,Martinez2020}. With the new developments, the range of applications increase, due to e.g. the new TMD merging method, which allows a description of the whole $p_\bot$-spectrum of the Drell Yan process and an accurate description of jets, even at high mltiplicity.\\
The first PB TMD PDF set within the four-flavor schemes along with the already existing sets in the five-flavor scheme opens the further investigation of PB evolution and PB TMD showers.\\
Other developments of the PB TMDs will lead to an increase in precision or an extension of the kinematical range. The first inclusion QED effects in the Parton Branching method, including the first photon TMD within the method is obtained. The inclusion of TMD splitting functions~\cite{CATANI1994475,Hautmann:2012sh,Gituliar2016,Hentschinski:2016wya,Hentschinski2018,Hentschinski:2021lsh} is underway~\cite{keersmaekers2021implementing}, and 
it is a first step towards a Monte Carlo that incorporates small-$x$ dynamics.

\section*{Acknowledgements}
We thank F. Hautmann, M. Hentschinski, H. Jung, A. Bermudez Martinez, A. Kusina, K. Kutak and A. Lelek for collaboration and discussion.  STM thanks the Humboldt
Foundation for the Georg Forster research fellowship.
%\begin{thebibliography}{99}
%\bibitem{...}
%....
%
%\end{thebibliography}
\bibliography{EPS}

\end{document}